\documentclass [6pt,a4paper]{article}
\usepackage{bbm}
\usepackage{amsfonts}
\usepackage{mathrsfs}
\usepackage {amssymb}
\usepackage {amsmath}
\usepackage{amsthm}
\usepackage{latexsym}
\usepackage{booktabs}
\usepackage{mathrsfs}
\usepackage{bm}
\usepackage{booktabs}

\def\dse#1{\vskip 0.6cm\noindent
        {\large\bf #1}
        \vskip 0.4cm}

\def\dse#1{\vskip 0.6cm\noindent
        {\large\bf #1}
        \vskip 0.4cm}
\usepackage{lastpage}
\usepackage{epsfig}

\begin{document}
\begin{center}
\textbf{\large{Entanglement-assisted quantum MDS codes constructed from constacyclic codes}}\footnote {addresses: School of Mathematics, Hefei University
of
Technology, Hefei 230009, Anhui, P.R. China.\\
E-mail: chenxiaojing0909@126.com (X.Chen), zhushixin@hfut.edu.cn (S.Zhu),  kxs6@sina.com (X.Kai).\\
 This research is supported by the National Natural Science
Foundation of China ( No.61772168;\ No.61572168).}
\end{center}

\begin{center}
{ { Xiaojing Chen, \  Shixin Zhu, \ Xiaoshan Kai} }
\end{center}

\begin{center}
\textit{School of Mathematics, Hefei University of
Technology, Hefei 230009, Anhui, P.R.China }
\end{center}

\noindent\textbf{Abstract:} Recently, entanglement-assisted quantum error correcting codes (EAQECCs) have been constructed by cyclic codes and negacyclic codes. In this paper, by analyzing the cyclotomic cosets in the defining set of constacyclic codes, we constructed three classes of new EAQECCs which satisfy the entanglement-assisted quantum Singleton bound. Besides, three classes of EAQECCs with maximal entanglement from constacyclic codes are constructed in the meanwhile.\\


\dse{1~~Introduction} Since the significant discovery in \cite{ref1} and \cite{ref2}, the theory of quantum
error-correcting codes (QECCs) has experienced tremendous growth. Many good QECCs have been constructed by using classical error-correcting codes \cite{ref3}-\cite{ref8}. In kinds of methods of constructing QECCs, the CSS construction is the most important one which provides stabilizer codes by exploiting the link between classical and quantum codes. However, the condition of dual-containing forms a barrier in the development of quantum coding theory. This problem get solved after Brun
et al. \cite{ref9} proposed the EAQECCs. In their paper, they proved that non-dual-containing classical quaternary codes can be used to construct EAQECCs if the sender and receiver shared entanglement in advance. EAQECCs allow the use of arbitrary classical codes (not necessarily self-orthogonal) for quantum data transmission via pre-shared entanglement bits. This inspire more and more researchers to focus on constructing good EAQECCs \cite{ref10}-\cite{ref15}.

Customarily, an entanglement-assisted quantum error correcting code (EAQECC) can be denoted as $[[n,k,d;c]]_q$, which encodes $k$ information qubits into $n$
channel qubits with the help of $c$ pairs of maximally entangled states and corrects up to
$\lfloor\frac{d-1}{2}\rfloor$ errors, where $d$ is the minimum distance of the code.  If $n-k=c$, the code is called an EAQECC with maximal entanglement. The performance of an EAQECC is measured by its rate and net rate $\frac{k-c}{n}$. When the net rate of an EAQECC is positive it is possible to obtain catalytic codes as shown by Brun et al. \cite{ref26}. Li et al. \cite{ref16} proposed the concept about a decomposition of the defining set of BCH
cyclic codes, transformed the problem of calculating the number of share pairs into determining a special subset of the defining set of a BCH code, and constructed some EAQECCs with good parameters. In Refs. \cite{ref17}, L$\ddot{u}$ and Li made a further study on constructing of
EAQECCs by using primitive quaternary BCH codes. Recently,  Chen et al. \cite{ref18}
generalize their method to apply in negacyclic codes, and
obtain four classes of optimal EAQECCs and two classes of maximal entanglement entanglement-assisted quantum codes. Lu et al. \cite{ref25} constructed six classes of $q$-qry entanglement-assisted quantum MDS codes based on classical negacyclic MDS codes.

As we all known, there exist optimal symmetric and asymmetric quantum codes of length $n=\frac{q^2+1}{5}$, where $q$ is some prime power. In Refs. \cite{ref19}, Kai et al. obtained the quantum MDS codes from constacyclic codes where $n=\frac{q^2+1}{5}$ and $q$ is an odd prime power. Afterwards, Chen et al. \cite{ref20}
constructed some asymmetric quantum MDS codes by using constacyclic codes where $n=\frac{q^2+1}{5}$ and $q$ is an even prime power.
Inspired by the above work, we consider constructing EAQECCs by constacyclic codes naturally. In this paper, we obtain three classes of new optimal EAQECCs by
constacyclic codes. Speaking specifically, for an even prime power $q=2^e$, where $e$ is odd, and an odd prime power $q$ with the form $20m+3$ or $20m+7$, we construct three classes of EAQECCs with parameters as follows:

\begin{itemize}
\item[(1)] $[[\frac{q^2+1}{5},\frac{q^2-6q+33}{5}-4t,\frac{3q-1}{5}+2t;4]]_q$, where $e\equiv1 \bmod\ 4$ and $1\leq t\leq\frac{q+3}{5}$.
\item[(2)] $[[\frac{q^2+1}{5},\frac{q^2-6q+29}{5}-4t,\frac{3q+1}{5}+2t;4]]_q$, where $e\equiv3 \bmod\ 4$ and $1\leq t\leq\frac{q+2}{5}$.
\item[(3)] $[[\frac{q^2+1}{5},\frac{q^2+1}{5}-q-4t+1,\frac{q+1}{2}+2t+2;4]]_q$, where $q$ is an odd prime power with the form $20m+3$ or $20m+7$, $m$ is a positive integer and $m\leq t\leq\frac{q-3}{4}$.
\end{itemize}

We also obtain three classes of maximal-entanglement entanglement-assisted quantum codes with parameters as follows:

\begin{itemize}
	\item[(1)] $[[\frac{q^2+1}{5},\frac{q^2+1}{5}-4,d\geq2;4]]_q$, where $e\equiv1 \bmod\ 4$.
	\item[(2)] $[[\frac{q^2+1}{5},\frac{q^2+1}{5}-4,d\geq2;4]]_q$, where $e\equiv3 \bmod\ 4$.
	\item[(3)] $[[\frac{q^2+1}{5},\frac{q^2+1}{5}-4,d\geq2;4]]_q$, where $q$ is an odd prime power with the form $20m+3$ or $20m+7$, $m$ is a positive integer.
\end{itemize}

This paper is organized as follows. In Section 2, some basic
background and results about constacyclic codes are reviewed. In
Section 3, we briefly review some basic definitions and results of EAQECCs. In Section 4, we construct three classes of optimal EAQECCs and three classes of maximal-entanglement entanglement-assisted quantum codes. Section 5 concludes the paper.

\dse{2~~Preliminaries}
 Let $\mathbb{F}_{q^{2}}$ be a finite field with $q^{2}$ elements,
 where $q$ is a power of a prime $p$. For any element $x\in\mathbb{F}_{q^{2}}$,
 we denote the conjugate $x^{q}$ of $x$ by $\overline{x}$. Given two vectors
 $\mathbf{x}=(x_0,x_1,\ldots,x_{n-1})$ and $\mathbf{y}=(y_0,y_1,\ldots,y_{n-1})\in \mathbb{F}_{q^{2}}^{n}$,
 their Hermitian inner product is defined as
  \[ \langle \mathbf{x},\mathbf{y}\rangle=x_0\overline{y}_0+x_1\overline{y}_1+\cdots+x_{n-1}\overline{y}_{n-1}\in\mathbb{F}_{q^{2}}.\]
 The vectors $\mathbf{x}$ and $\mathbf{y}$ are called orthogonal with respect to the Hermitian
 inner product if $\langle \mathbf{x},\mathbf{y}\rangle=0$.
 A $q^{2}$-ary linear code  $\mathcal{C}$ of length $n$ is a nonempty
 subspace of the vector space $\mathbb{F}_{q^{2}}^{n}$.
 For a $q^{2}$-ary linear code  $\mathcal{C}$, the Hermitian
 dual code of $\mathcal{C}$ is defined as
 \[\mathcal{C}^{\bot_{h}}=\{\mathbf{x}\in\mathbb{F}_{q^{2}}^{n}|~\langle \mathbf{x},\mathbf{y}\rangle=0 ~for~ all~\mathbf{y}\in\mathcal{C}\}.\]
 A $q^{2}$-ary linear code $\mathcal{C}$ of length $n$ is called Hermitian
 self-orthogonal if $\mathcal{C}\subseteq\mathcal{C}^{\bot_{h}}$, and it is called Hermitian self-dual if
  $\mathcal{C}=\mathcal{C}^{\bot_{h}}$.
 For a nonzero element $\lambda$ of $\mathbb{F}_{q^{2}}$, if $\mathcal{C}$ is
 closed under the $\lambda$-constacyclic shift, i.e., if
 $(x_0,x_1,\ldots,x_{n-1})\in\mathcal{C}$ implies $(\lambda x_{n-1},x_0,\ldots,x_{n-2})\in\mathcal{C}$,
 then $\mathcal{C}$ is said to be a $\lambda$-constacyclic code.
 Customarily, a codeword $\textbf{c}=(c_0,c_1,\ldots,c_{n-1})$ in $\mathcal{C}$
 is identified with its polynomial representation $c(x)=c_0+c_1x+\cdots+c_{n-1}x^{n-1}$.
 It is well known that a $\lambda$-constacyclic code $\mathcal{C}\in\mathbb{F}_{q^{2}}^{n}$
 is an ideal of the quotient ring $\mathbb{F}_{q^{2}}[x]/\langle x^{n}-\lambda\rangle$
 and $\mathcal{C}$ can be generated by a monic divisor $g(x)$
 of $x^{n}-\lambda$. The polynomial $g(x)$ is called the generator
 polynomial of the code $\mathcal{C}$ and the dimension of $\mathcal{C}$
 is $n-k$, where $k=\deg(g(x))$.

In the following, we take $q$ is a power of a prime $p$, $r$ is the order of $\lambda$ in  $\mathbb{F}^*_{q^{2}}$. Note that $\lambda\bar{\lambda}=1$ in $\mathbb{F}_{q^{2}}$. Assume that $\gcd(q,n)=1$. Let $\delta$ be a primitive $rn$-th root of unity in some extension field of
 $\mathbb{F}_{q^{2}}$ such that $\delta^n=\eta$. Let $\xi=\delta^r$, then $\xi$ is a primitive
 $n$-th root of unity. Hence,
 \[x^{n}-\lambda=\prod_{j=0}^{n-1}(x-\delta\xi^j)=\prod_{j=0}^{n-1}(x-\delta^{1+jr}).\]

 Let $\Omega=\{1+jr|~0\leq j \leq n-1\}$. For each $i\in\Omega$, let $\mathbb{C}_i$ be the
 $q^2$-cyclotomic coset modulo $rn$ containing $i$,
 \[\mathbb{C}_i=\{i,iq^2,iq^4,\ldots,iq^{2(m_i-1)}\},\]
 where $m_i$ is the smallest positive integer such that $iq^{2m_i}\equiv i \mod rn$. Each $\mathbb{C}_i$
 corresponds to an irreducible divisor of  $x^{n}-\lambda$ over $\mathbb{F}_{q^{2}}$. Let $\mathcal{C}$
 be an $\lambda$-constacyclic code of length $n$ over $\mathbb{F}_{q^{2}}$ with generator polynomial
 $g(x)$. Then the set $Z=\{i\in\Omega|~g(\delta^i)=0\}$ is called the defining set of $\mathcal{C}$.
 Obviously, the defining set of $\mathcal{C}$ must be a union of some  $q^2$-cyclotomic cosets modulo $rn$ and
 dim$(\mathcal{C})=n-|Z|$. It is clear to see that $\mathcal{C}^{\bot_{h}}$ has defining set 
$Z^{\bot_h}=\{z\in\Omega|~-qz~\textrm{mod}~rn~\notin Z\}$.
Note that $Z^{-q}=\{-qz~\textrm{mod}~rn|~z\in Z\}$. Then $\mathcal{C}$ contains its Hermitian dual
 code if and only if $Z\cap Z^{-q}=\emptyset$ from lemma 2.2 in Refs. \cite{ref19}.\\

 Similar to cyclic codes, there exists the following BCH bound for
 $\lambda$-constacyclic codes in Refs. \cite{ref21} and \cite{ref22}.\\

\noindent\textbf{Theorem 2.1 (The BCH bound for constacyclic codes)} \emph{ Assume that $gcd(q,n)=1$.
Let $\mathcal{C}$ be an $\lambda$-constacyclic code of length $n$ over  $\mathbb{F}_{q^{2}}$, and let its
generator polynomial $g(x)$ have the elements $\{\delta^{1+jr}|~0\leq j\leq d-2\}$ as the roots, where
$\delta$ is a primitive $rn$-th root of unity. Then the minimum distance of $\mathcal{C}$ is at least $d$.}
\\

\dse{3~~Review of EAQECCs }  In this section, we give some basic definitions and results of EAQECCs. More details about EAQECCs theory, please refer to Refs. \cite{ref9}-\cite{ref18} therein.

Suppose that $H$ is an $(n-k)\times n$ parity check matrix of $\mathcal{C}$ over $\mathbb{F}_{q^{2}}$. Then, $\mathcal{C}^{\bot_{h}}$ has an $n\times (n-k)$ generator matrix $H^\dagger$, where $H^\dagger$ is the conjugate transpose matrix of $H$ over $\mathbb{F}_{q^{2}}$.

The following proposition is about the Singleton bound of classical linear codes in Ref.\cite{ref23}.\\

\noindent\textbf{Lemma 3.1 (Singleton bound)} \emph {~If an $[n,k,d]$ linear code over $\mathbb{F}_{q}$
	exists, then $k\leq n-d+1$. If the equality $k=n-d+1$ holds, then the code is an MDS code.
}\\

In the following, we recall several results which are important for constructing EAQECCs in Refs. \cite{ref9}, \cite{ref10} and \cite{ref13}.\\

\noindent\textbf{Theorem 3.2} [9,10]\emph{ If
	$\mathcal{C}=[n,k,d]_{q^2}$ is a classical code  over $\mathbb{F}_{q^2}$ and
	$H$ is its parity check matrix, then $\mathcal{C}^{\bot_{h}}$ EA stabilizes an  entanglement-assisted code with parameters $[[n,2k-n+c,d;c]]_{q}$, where $c=rank(HH^\dag)$ is the number of maximally entangled states required and $H^\dag$ is the conjugate matrix of $H$ over $\mathbb{F}_{q^2}$. }\\

\noindent\textbf{Theorem 3.3} [9,14]\emph{ Assume that
	$\mathcal{C}=[n,k,d;c]_{q}$ is an entanglement-assisted quantum code, where	$d\leq \frac{n+2}{2}$, then $\mathcal{C}$ satisfies the entanglement-assisted Singleton bound $n+c-k\geq 2(d-1)$. If $\mathcal{C}$ satisfies the equality $n+c-k= 2(d-1)$ for $d\leq \frac{n+2}{2}$, then it is called an entanglement-assisted quantum MDS code. }\\

\dse{4 ~~Construction of entanglement-assisted quantum MDS codes} In Refs.\cite{ref16}-\cite{ref25}, the authors gave definitions for decomposing the defining set of cyclic codes and negacyclic codes. We define a decomposion of the defining set of constacyclic codes as follows.\\

\noindent\textbf{Definition 4.1} \emph { Let $C$ be a
	constacyclic code of length $n$ with defining set $Z$.
	Assume that $Z_1=Z\cap(-qZ)$ and $Z_2=Z\setminus Z_1$, where $-qZ=\{n-qx|x\in Z\}$. Then, $Z=Z_1\cup Z_2$ is called a decomposion of the defining set of $C$.}\\

In the following, we give a lemma which is a generalization of Lemma 1 in Refs.\cite{ref18}. The proof is similar, so we omit it there.\\

\noindent\textbf{Lemma 4.2} \emph { Let $C$ be a constacyclic code of length $n$ over $\mathbb{F}_{q^{2}}$, where $gcd(n,q)=1$. Suppose that $Z$ is the defining set of the constacyclic code $C$ and $Z=Z_1\cup Z_2$ is a decomposition of $Z$. Then, the number of entangled states requires is $C=|Z_1|$.}\\

\noindent\textbf{Lemma 4.3 }\cite{ref20} \emph { Let $q=2^e$, where $e>1$
	is odd, $n=\frac{q^2+1}{5}$, $s=\frac{(q+6)n}{2}$, $r=\frac{q^2-q}{2}$, where $r=s-\frac{(q+1)(n+1)}{2}$ and $\Omega=\{1+(q+1)j|~0\leq j\leq n-1\}$. Then, for any integer $i\in\Omega$,
	the $q^2$-cyclotomic cosets $\mathbb{C}_i$ modulo $(q+1)n$ is given by
	\begin{itemize}
	\item [1)] ~$\mathbb{C}_s=\{s\}$ and  $\mathbb{C}_{s-(q+1)j}=\{s-(q+1)j,s+(q+1)j\}$ for $1\leq j\leq\frac{n-1}{2}$.
	\item [2)] ~Let $q=2^{e}$ with
	$e\equiv1 \bmod\ 4$. If $\mathcal C$ is an $\lambda$-constacyclic
	code over $\mathbb{F}_{q^2}$ of length $n$
	with defining set $Z=\bigcup_{j=0}^\delta \mathbb{C}_{r-(q+1)j}$, where $0\leq\delta\leq\frac{3q-16}{10}$,
	then $\mathcal{C}^{\bot_{h}}\subseteq\mathcal{C}$.
	\item [3)] ~Let $q=2^{e}$ with
	$e\equiv3\bmod\ 4$. If $\mathcal C$ is an $\lambda$-constacyclic
	code over $\mathbb{F}_{q^2}$ of length $n$
	with defining set $Z=\bigcup_{j=0}^\delta \mathbb{C}_{r-(q+1)j}$, where $0\leq\delta\leq\frac{3q-14}{10}$,
	then $\mathcal{C}^{\bot_{h}}\subseteq\mathcal{C}$.	\end{itemize}}

Based on the discussions above, we give the first important theorem of this paper below.\\

\noindent\textbf{Theorem 4.4 }\emph{Let $q=2^{e}$ with
	$e\equiv1(\bmod\ 4)$. Let $n=\frac{q^2+1}{5}$, $s=\frac{(q+6)n}{2}$ and
	$r=\frac{q^2-q}{2}$, where $r=s-\frac{(q+1)(n+1)}{2}$. If $\mathcal C$ is a $q^2$-ary constacyclic code of length $n$
	with defining set $Z=\bigcup_{i=0}^{\frac{3q-16}{10}+t} \mathbb{C}_{r-(q+1)i}$, where $1\leq t\leq\frac{q+3}{5}$,
	then there exsit EAQECCs with parameters $[[\frac{q^2+1}{5},\frac{q^2-6q+33}{5}-4t,\frac{3q-1}{5}+2t;4]]_q$.}\\

\noindent\textbf{Proof.} From Lemma 4.3, we can assume that the defining set of the constacyclic code $\mathcal C$ is $Z=\bigcup_{i=0}^{\frac{3q-16}{10}+t} \mathbb{C}_{r-(q+1)i}$, where $1\leq t\leq\frac{q+3}{5}$. Then  $\mathcal C$ is a constacyclic code with parameters $[\frac{q^2+1}{5},\frac{q^2+1}{5}-\frac{3q-16}{5}-2t-2,\frac{3q-16}{5}+2t+3]_{q^2}$ from Theorem $2.1$ and Lemma $3.1$. Therefore, we have the following result.\\

\begin{align}
Z_1&=Z\cap (-qZ) \notag \\
&=((\cup_{i=0}^{\frac{3q-16}{10}}\mathbb{C}_{a})\cup(\cup_{i=\frac{3q-6}{10}}^{\frac{3q-16}{10}+t}\mathbb{C}_{a}))\notag\\
&\cap(-q(\cup_{i=0}^{\frac{3q-16}{10}}\mathbb{C}_{a})\cup-q(\cup_{i=\frac{3q-6}{10}}^{\frac{3q-16}{10}+t}\mathbb{C}_{a}))\notag\\
&=((\cup_{i=0}^{\frac{3q-16}{10}}\mathbb{C}_{a})\cap-q(\cup_{i=0}^{\frac{3q-16}{10}}\mathbb{C}_{a}))\notag\\
&\cup((\cup_{i=0}^{\frac{3q-16}{10}}\mathbb{C}_{a})\cap-q(\cup_{i=\frac{3q-6}{10}}^{\frac{3q-16}{10}+t}\mathbb{C}_{a}))\notag\\
&\cup((\cup_{i=\frac{3q-6}{10}}^{\frac{3q-16}{10}+t}\mathbb{C}_{a})\cap-q(\cup_{i=0}^{\frac{3q-16}{10}}\mathbb{C}_{a}))\notag\\
&\cup((\cup_{i=\frac{3q-6}{10}}^{\frac{3q-16}{10}+t}\mathbb{C}_{a})\cap-q(\cup_{i=\frac{3q-6}{10}}^{\frac{3q-16}{10}+t}\mathbb{C}_{a}))\notag\\
&=\mathbb{C}_\frac{q^2-q+3}{5}\cup\mathbb{C}_\frac{2q^2-2q+1}
{5},
\end{align}
where $a=r-(q+1)i$.

From Lemma 4.3, we have  \[(\cup_{i=0}^{\frac{3q-16}{10}}\mathbb{C}_{a})\cap-q(\cup_{i=0}^{\frac{3q-16}{10}}\mathbb{C}_{a})=\emptyset.\]
In order to get the result of equation $(1)$, we have to show that
\begin{align*}
(\cup_{i=\frac{3q-6}{10}}^{\frac{3q-16}{10}+t}\mathbb{C}_{a})\cap-q(\cup_{i=0}^{\frac{3q-16}{10}}\mathbb{C}_{a})&=\mathbb{C}_\frac{q^2-q+3}{5},\\
(\cup_{i=0}^{\frac{3q-16}{10}}\mathbb{C}_{a})\cap-q(\cup_{i=\frac{3q-6}{10}}^{\frac{3q-16}{10}+t}\mathbb{C}_{a})&=\mathbb{C}_\frac{2q^2-2q+1}{5},
\end{align*}
and\\
\[(\cup_{i=\frac{3q-6}{10}}^{\frac{3q-16}{10}+t}\mathbb{C}_{a})\cap-q(\cup_{i=\frac{3q-6}{10}}^{\frac{3q-16}{10}+t}\mathbb{C}_{a})=\emptyset.\]

Firstly, we show that
\[(\cup_{i=\frac{3q-6}{10}}^{\frac{3q-16}{10}+t}\mathbb{C}_{a})\cap-q(\cup_{i=0}^{\frac{3q-16}{10}}\mathbb{C}_{a})=\mathbb{C}_\frac{q^2-q+3}{5}.\]
It is easy to show that $-q\mathbb{C}_\frac{q^2-q+3}{5}=\mathbb{C}_\frac{2q^2-2q+1}{5}$. Therefore, we have

\begin{align*}
&(\cup_{i=\frac{3q-6}{10}}^{\frac{3q-16}{10}+t}\mathbb{C}_{a})\cap-q(\cup_{i=0}^{\frac{3q-16}{10}}\mathbb{C}_{a})\\
&=(\mathbb{C}_\frac{q^2-q+3}{5}\cup(\cup_{i=\frac{3q+4}{10}}^
{\frac{3q-16}{10}+t}\mathbb{C}_{a}))
\cap
-q(\cup_{i=0}^{\frac{3q-16}{10}}\mathbb{C}_{a})
\\
&=(\mathbb{C}_\frac{q^2-q+3}{5}\cap-q(\cup_{i=0}^{\frac{3q-16}{10}}\mathbb{C}_{a}))\\
&\cup((\cup_{i=\frac{3q+4}{10}}^
{\frac{3q-16}{10}+t}\mathbb{C}_{a})\cap
-q(\cup_{i=0}^{\frac{3q-16}{10}}\mathbb{C}_{a}))\\
&=\mathbb{C}_\frac{q^2-q+3}{5}
\end{align*}

In fact,
\[\mathbb{C}_\frac{q^2-q+3}{5}\cap-q(\cup_{i=0}^{\frac{3q-16}{10}}\mathbb{C}_{a})=\mathbb{C}_\frac{q^2-q+3}{5}\]
from Lemma 4.3 and
\[(\cup_{i=\frac{3q+4}{10}}^
{\frac{3q-16}{10}+t}\mathbb{C}_{a})\cap
-q(\cup_{i=0}^{\frac{3q-16}{10}}\mathbb{C}_{a})=\emptyset,\]
for $2\leq t\leq\frac{q+3}{5}$.\\

If $(\cup_{i=\frac{3q+4}{10}}^
{\frac{3q-16}{10}+t}\mathbb{C}_{a})\cap
-q(\cup_{i=0}^{\frac{3q-16}{10}}\mathbb{C}_{a})\neq\emptyset$
for $2\leq t\leq\frac{q+3}{5}$, i.e.,\\
\[(\cup_{i=2}^
{t}\mathbb{C}_{r-(q+1)(i+\frac{3q-16}{10})})\cap
-q(\cup_{i=0}^{\frac{3q-16}{10}}\mathbb{C}_{r-(q+1)i})\neq\emptyset\]
for  $2\leq t\leq\frac{q+3}{5}$,
then there exist two integers $l$ and $j$, where $2\leq l\leq \frac{q+3}{5},~0\leq j\leq\frac{3q-16}{10}$,
such that \[r-(q+1)(l+\frac{3q-16}{10})\equiv-q[r-(q+1)j]q^{2k} \bmod\ (q+1)n\]
for some $k\in\{0,1\}$. We can seek contradictions as follows.

\begin{itemize}
	\item [(i)]  When $k=0$, we have
	$r-(q+1)(l+\frac{3q-16}{10})\equiv-q[r-(q+1)j] \bmod\ (q+1)n$.
	Since $(q+1)r \equiv (q+1)(l+\frac{3q-16}{10}+qj) \bmod\ (q+1)n$, we have $r\equiv l+\frac{3q-16}{10}+qj \bmod\ n$, which is equal to
	\begin{align} \frac{2q+6}{5}+q\cdot\frac{q-12}{10}\equiv l+qj \bmod\ n,
	\end{align}
	where $n=\frac{q^2+1}{5}=\frac{2q+1}{5}+q\cdot \frac{q-2}{5}$. From
	$ 2\leq l\leq\frac{q+3}{5}<\frac{2q+1}{5}$,	we have the following results.
 	\item [(a)] when $0\leq j\leq\frac{q-2}{5}$, $l+qj<n$.
	If equation (2) establish, we have $l=\frac{2q+6}{5}$, $j=\frac{q-12}{10}$. Because $l_{max}=\frac{q+3}{5}<\frac{2q+6}{5}$, it is a contradiction.
	\item [(b)] when $\frac{q-2}{5}<j\leq\frac{3q-16}{10}$, from $j$ is an integer, we have $\frac{q+3}{5}\leq j\leq\frac{3q-16}{10}$, so $l+qj\geq qj\geq q\cdot\frac{q+3}{5}=\frac{q^2+3q}{5}>n$. From $2\leq l\leq\frac{q+3}{5}, \frac{q+3}{5}\leq j\leq\frac{3q-16}{10}$, we can get
	$ 2+q\cdot\frac{q+3}{5}\leq l+qj\leq\frac{q+3}{5}+q\cdot\frac{3q-16}{10}$, i.e.,
	$\frac{3q+9}{5}\leq l+qj-n\leq\frac{q^2-14q+4}{10}<n$. Because
	$\frac{q^2-14q+4}{10}<\frac{q^2-8q+12}{10}=\frac{2q+6}{5}+q\cdot\frac{q-12}{10}$, equation $(2)$ is not establish.
\end{itemize}

\begin{itemize}
	\item [(ii)]   When $k=1$, we have
	$r-(q+1)(l+\frac{3q-16}{10})\equiv-q^3[r-(q+1)j] \bmod\ (q+1)n$, which is
	equal to
	\begin{align}
	\frac{11q-2}{10}\equiv j+ql \bmod\ n,
	\end{align}
	where $n=\frac{q^2+1}{5}=\frac{2q+1}{5}+q\cdot\frac{q-2}{5}$.
	From $ 0\leq j\leq\frac{3q-16}{10}<\frac{2q+1}{5}$, we
	have the following results.
	\item [(a)] when $2\leq l\leq\frac{q-2}{5}$, $j+ql<n$.
	If equation (3) establish, we have $j=\frac{11q-2}{10}$, $l=0$. Because $l_{min}=2>0$, it is a contradiction.
	\item [(b)]  when $\frac{q-2}{5}< l\leq\frac{q+3}{5}$, because $l$ is an integer, we have $l=\frac{q+3}{5}$, so $j+ql\geq ql= \frac{q^2+3q}{5}>n$. From
$l=\frac{q+3}{5}, 0\leq j\leq\frac{3q-16}{10}$, we have
	$q\cdot\frac{q+3}{5}\leq j+ql\leq\frac{3q-16}{10}+q\cdot\frac{q+3}{5}$,
    i.e.,
	$\frac{3q-1}{5}\leq j+ql-n\leq\frac{9q-18}{10}<n$, because
	$ \frac{9q-18}{10}<\frac{11q-2}{10}$, therefore equation
	$(3)$ is not establish.
\end{itemize}

From the above discussions, we can see\\
\[(\cup_{i=\frac{3q-6}{10}}^
{\frac{3q-16}{10}+t}\mathbb{C}_{a})\cap
-q(\cup_{i=0}^{\frac{3q-16}{10}}\mathbb{C}_{a})=\mathbb{C}_{\frac{q^2-q+3}{5}},\] for $1\leq t\leq\frac{q+3}{5}$.\\

Secondly, we show that \\
\[(\cup_{i=0}^
{\frac{3q-16}{10}}\mathbb{C}_{a})\cap
-q(\cup_{i=\frac{3q-6}{10}}^{\frac{3q-16}{10}+t}\mathbb{C}_{a})=\mathbb{C}_{\frac{2q^2-2q+1}{5}}.\]
Since\\
\[-q((\cup_{i=\frac{3q-6}{10}}^
{\frac{3q-16}{10}+t}\mathbb{C}_{a})\cap
-q(\cup_{i=0}^{\frac{3q-16}{10}}\mathbb{C}_{a}))=-q\mathbb{C}_{\frac{q^2-q+3}{5}}=\mathbb{C}_{\frac{2q^2-2q+1}{5}},\]
it follows that \\
 \[(\cup_{i=0}^
{\frac{3q-16}{10}}\mathbb{C}_{a})\cap
-q(\cup_{i=\frac{3q-6}{10}}^{\frac{3q-16}{10}+t}\mathbb{C}_{a})=\mathbb{C}_{\frac{2q^2-2q+1}{5}}.\]

Finally, we show that\\
\[(\cup_{i=\frac{3q-6}{10}}^
{\frac{3q-16}{10}+t}\mathbb{C}_{a})\cap
-q(\cup_{i=\frac{3q-6}{10}}^{\frac{3q-16}{10}+t}\mathbb{C}_{a})=\emptyset,\]
for $1\leq t\leq\frac{q+3}{5}$.
If $(\cup_{i=\frac{3q-6}{10}}^
{\frac{3q-16}{10}+t}\mathbb{C}_{a})\cap
-q(\cup_{i=\frac{3q-6}{10}}^{\frac{3q-16}{10}+t}\mathbb{C}_{a})\neq\emptyset,$, i.e.,
\[(\cup_{i=1}^{t}\mathbb{C}_{r-(q+1)(i+\frac{3q-16}{10})})\cap-q(\cup_{i=1}^{t}\mathbb{C}_{r-(q+1)(i+\frac{3q-16}{10})})\neq
\emptyset\]
 for $1\leq t \leq \frac{q+3}{5}$, then there exist two integers $l$ and $j$, where $1\leq l,j \leq \frac{q+3}{5}$, such that\\
\[r-(q+1)(l+\frac{3q-16}{10})\equiv-q[r-(q+1)(j+\frac{3q-16}{10})]q^{2k} \bmod\ (q+1)n,\]
 for some
$k\in\{0,1\}$. We can seek contradictions as follows.

\begin{itemize}
	\item [(i)]   When $k=0$, we have
	$r-(q+1)(l+\frac{3q-16}{10})\equiv-q[r-(q+1)(j+\frac{3q-16}{10})] \bmod\ (q+1)n$, which is equal to
   \begin{align}
   \frac{4q+7}{5}\equiv l+qj \bmod\ n,
   \end{align}
   where $n=\frac{q^2+1}{5}=\frac{2q+1}{5}+q\cdot \frac{q-2}{5}$.
	From $ 1\leq l\leq\frac{q+3}{5}<\frac{2q+1}{5}$, we have the following results.
	\item [(a)] when $1\leq j\leq\frac{q-2}{5}$, $l+qj<n$.
	If equation (4) establish, we have $l=\frac{4q+7}{5}$, $j=0$. Because $j_{min}=1>0$,  it is a contradiction.
	\item [(b)] when $\frac{q-2}{5}< j\leq\frac{q+3}{5}$, because $j$ is an integer, we have $j=\frac{q+3}{5}$, so $l+qj\geq qj= \frac{q^2+3q}{5}>n$. From $1\leq l\leq\frac{q+3}{5},  j=\frac{q+3}{5}$, we can get
	$1+q\cdot\frac{q+3}{5}\leq l+qj\leq\frac{q+3}{5}+q\cdot\frac{q+3}{5}$, i.e.,
	$\frac{3q+4}{5}\leq l+qj-n\leq\frac{4q+2}{5}<n$. Because
	$\frac{4q+2}{5}<\frac{4q+7}{5}$, equation
	$(4)$ is not establish.\\
\end{itemize}

\begin{itemize}
	\item [(ii)] When $k=1$, we have
	$r-(q+1)(l+\frac{3q-16}{10})\equiv-q^3[r-(q+1)(j+\frac{3q-16}{10})] \bmod\ (q+1)n$, which
    is equal to
    \begin{align}
    \frac{4q+7}{5}\equiv j+ql \bmod\ n ,
    \end{align}
	where $n=\frac{q^2+1}{5}=\frac{2q+1}{5}+q\cdot\frac{q-2}{5}$.
	From $1\leq j\leq\frac{q+3}{5}<\frac{2q+1}{5}$, we have the following results.
	\item [(a)]  when $1\leq l\leq\frac{q-2}{5}$, $l+qj<n$.
	If equation (5) establish, we have $j=\frac{4q+7}{5}$, $l=0$. But $l_{min}=1>0$, it is a contradiction.
	\item [(b)]  when $\frac{q-2}{5}< l\leq\frac{q+3}{5}$, because $l$ is an integer, we have $l=\frac{q+3}{5}$, so $j+ql\geq ql= \frac{q^2+3q}{5}>n$. From $1\leq j\leq\frac{q+3}{5},  l=\frac{q+3}{5}$, we can get
	$ 1+q\cdot\frac{q+3}{5}\leq j+ql\leq\frac{q+3}{5}+q\cdot\frac{q+3}{5}$, i.e.,
	$\frac{3q+4}{5}\leq j+ql-n\leq\frac{4q+2}{5}$. Because
	$\frac{4q+2}{5}<\frac{4q+7}{5}$, equation
	$(5)$ is not establish. It follows that $(\cup_{i=\frac{3q-6}{10}}^{\frac{3q-16}{10}+t}\mathbb{C}_{e})\cap(-q\cup_{i=\frac{3q-6}{10}}^{\frac{3q-16}{10}+t}\mathbb{C}_{e})=\emptyset$.
\end{itemize}

From Lemma 4.2, we have $c=4$. From Theorem 3.2, there exist entanglement assisted quantum codes with parameters $[[\frac{q^2+1}{5},\frac{q^2-6q+33}{5}-4t,\frac{3q-1}{5}+2t;4]]_q$, where $1\leq t\leq\frac{q+3}{5}$.\\

For the case of $q=2^{e}$ with $e\equiv3 \bmod\ 4$, we can produce the following entanglement-assisted quantum MDS codes.
The proof is similar to that in the case of  $q=2^{e}$ with $e\equiv1 \bmod\ 4$ and we omit it there.\\

\noindent\textbf{Theorem 4.5 }\emph{Let $q=2^{e}$ with
	$e\equiv3 \bmod\ 4$. Let $n=\frac{q^2+1}{5}$, $s=\frac{(q+6)n}{2}$ and
	$r=\frac{q^2-q}{2}$,  where $r=s-\frac{(q+1)(n+1)}{2}$. If $\mathcal C$ is a $q^2$-ary constacyclic codes of length $n$
	with defining set $Z=\bigcup_{i=0}^{\frac{3q-14}{10}+t} \mathbb{C}_{r-(q+1)i}$, where $1\leq t\leq\frac{q+2}{5}$,
	then there exsit EAQECCs with parameters $[[\frac{q^2+1}{5},\frac{q^2-6q+29}{5}-4t,\frac{3q+1}{5}+2t;4]]_q$.}\\

\noindent\textbf{Lemma 4.6} \cite{ref19} \emph {  Let $n=\frac{q^2+1}{5}$, $s=\frac{q^2+1}{2}$. Then, for any integer $i\in\Omega=\{1+(q+1)j|~0\leq j\leq n-1\}$, the
 $q^2$-cyclotomic cosets $\mathbb{C}_i$ modulo $(q+1)n$ is given by
 	\begin{itemize}
 	\item [1)] $\mathbb{C}_s=\{s\}$ and $\mathbb{C}_{s+\frac{n(q+1)}{2}}=\{s+\frac{n(q+1)}{2}\}$, and ~$\mathbb{C}_{s-(q+1)j}=\{s-(q+1)j,s+(q+1)j\}$ for    $1\leq j\leq\frac{n}{2}-1$.
	\item [2)] Let $q$ be an odd prime power with the form $20m+3$ or $20m+7$, where $m$ is a positive integer. If $C$ is an $\omega^{q-1}$-constacyclic code over $\mathbb{F}_{q^{2}}$ of length $n$ with defining set $Z=\bigcup_{j=0}^{\delta}\mathbb{C}_{s-(q+1)j}$, where $0\leq \delta\leq
	\frac{q+1}{4}$, then  $\mathcal{C}^{\bot_{h}}\subseteq\mathcal{C}$.
	\end{itemize}}

From the discussions above, we can get the third theorem of this paper below.\\

\noindent\textbf{Theorem 4.7 }\emph{Let $n=\frac{q^2+1}{5}$ and $s=\frac{q^2+1}{2}$, where $q$ is an odd prime power with the form $20m+3$ or $20m+7$ and $m$ is a positive integer. If $\mathcal C$ is a $q^2$-ary constacyclic code of length $n$ with defining set  $Z=\bigcup_{i=0}^{\frac{q+1}{4}+t} \mathbb{C}_{s-(q+1)i}$, then there exist EAQECCs with parameters $[[\frac{q^2+1}{5},\frac{q^2+1}{5}-q-4t+1,\frac{q+1}{2}+2t+2;4]]_q$, where $m\leq t\leq\frac{q-3}{4}$.}\\

\noindent\textbf{Proof.} We only proof the case of $q$ is an odd prime power with the form $20m+3$. As for  $q$ is an odd prime power with the form $20m+7$, the proof is similar to that in the case of the former, and we omit it there. From Lemma 4.6, we can assume that the defining set of the constacyclic code $\mathcal C$ is $Z=\bigcup_{i=0}^{\frac{q+1}{4}+t} \mathbb{C}_{s-(q+1)i}$, then  $\mathcal C$ is a constacyclic code with parameters $[\frac{q^2+1}{5},\frac{q^2+1}{5}-\frac{q+1}{2}-2t-1,\frac{q+1}{2}+2t+2]_{q^2}$ from Theorem $2.1$ and Lemma $3.1$. Therefore, we have the following result.\\

\begin{align}
Z_1&=Z\cap (-qZ)\notag\\
&=((\cup_{i=0}^{\frac{q+1}{4}}\mathbb{C}_{b})\cup(\cup_{i=\frac{q+5}{4}}^{\frac{q+1}{4}+t}\mathbb{C}_{b}))\notag\\
&\cap
(-q(\cup_{i=0}^{\frac{q+1}{4}}\mathbb{C}_{b})\cup-q(\cup_{i=\frac{q+5}{4}}^{\frac{q+1}{4}+t}\mathbb{C}_{b}))\notag\\
&=((\cup_{i=0}^{\frac{q+1}{4}}\mathbb{C}_{b})\cap-q(\cup_{i=0}^{\frac{q+1}{4}}\mathbb{C}_{b}))\notag\\
&\cup((\cup_{i=0}^{\frac{q+1}{4}}\mathbb{C}_{b})\cap-q(\cup_{i=\frac{q+5}{4}}^{\frac{q+1}{4}+t}\mathbb{C}_{b}))\notag\\
&\cup((\cup_{i=\frac{q+5}{4}}^{\frac{q+1}{4}+t}\mathbb{C}_{b})\cap-q(\cup_{i=0}^{\frac{q+1}{4}}\mathbb{C}_{b}))\notag\\
&\cup((\cup_{i=\frac{q+5}{4}}^{\frac{q+1}{4}+t}\mathbb{C}_{b})\cap-q(\cup_{i=\frac{q+5}{4}}^{\frac{q+1}{4}+t}\mathbb{C}_{b}))\notag\\
&=\mathbb{C}_{\frac{(q-1)^2}{4}-m(q+1)}\cup\mathbb{C}_{s-2m(q+1)},
\end{align}
where $b=s-(q+1)i$.

From Lemma 4.6, we have  \[(\cup_{i=0}^{\frac{q+1}{4}}\mathbb{C}_{b})\cap-q(\cup_{i=0}^{\frac{q+1}{4}}\mathbb{C}_{b})=\emptyset.\]
In order to get the result of equation $(6)$, we have to show that
\begin{align*}
 (\cup_{i=\frac{q+5}{4}}^
{\frac{q+1}{4}+t}\mathbb{C}_{b})\cap
-q(\cup_{i=0}^{\frac{q+1}{4}}\mathbb{C}_{b})&=\mathbb{C}_{\frac{(q-1)^2}{4}-m(q+1)},\\
(\cup_{i=0}^{\frac{q+1}{4}}\mathbb{C}_{b})\cap-q(\cup_{i=\frac{q+5}{4}}^{\frac{q+1}{4}+t}\mathbb{C}_{b})&=\mathbb{C}_{s-2m(q+1)},\end{align*}
and\\
\[(\cup_{i=\frac{q+5}{4}}^{\frac{q+1}{4}+t}\mathbb{C}_{b})\cap-q(\cup_{i=\frac{q+5}{4}}^{\frac{q+1}{4}+t}\mathbb{C}_{b})=\emptyset.\]

Firstly, we show that
\[(\cup_{i=\frac{q+5}{4}}^
{\frac{q+1}{4}+t}\mathbb{C}_{b})\cap
-q(\cup_{i=0}^{\frac{q+1}{4}}\mathbb{C}_{b})=\mathbb{C}_{\frac{(q-1)^2}{4}-m(q+1)}.\]
It is easy to show that $-q\mathbb{C}_{\frac{(q-1)^2}{4}-m(q+1)}=\mathbb{C}_{s-2m(q+1)}$. Therefore, we have

\begin{align*}
&(\cup_{i=\frac{q+5}{4}}^
{\frac{q+1}{4}+t}\mathbb{C}_{b})\cap
-q(\cup_{i=0}^{\frac{q+1}{4}}\mathbb{C}_{b})\\
&=((\cup_{i=\frac{q+5}{4}}^
{\frac{q+1}{4}+m}\mathbb{C}_{b})\cup(\cup_{i=\frac{q+5}{4}+m}^
{\frac{q+1}{4}+t}\mathbb{C}_{b}))\cap
-q(\cup_{i=0}^{\frac{q+1}{4}}\mathbb{C}_{b})\\
&=(\cup_{i=\frac{q+5}{4}}^
{\frac{q+1}{4}+m}\mathbb{C}_{b}\cap-q(\cup_{i=0}^{\frac{q+1}{4}}\mathbb{C}_{b}))\\
&\cup((\cup_{i=\frac{q+5}{4}+m}^
{\frac{q+1}{4}+t}\mathbb{C}_{b})\cap
-q(\cup_{i=0}^{\frac{q+1}{4}}\mathbb{C}_{b}))\\
&=\mathbb{C}_{\frac{(q-1)^2}{4}-m(q+1)}.
\end{align*}
In fact,\\
\[(\cup_{i=\frac{q+5}{4}}^
{\frac{q+1}{4}+m}\mathbb{C}_{b})\cap-q(\cup_{i=0}^{\frac{q+1}{4}}\mathbb{C}_{b})=\mathbb{C}_{\frac{(q-1)^2}{4}-m(q+1)},\]
 and
\[(\cup_{i=\frac{q+5}{4}+m}^
{\frac{q+1}{4}+t}\mathbb{C}_{b})\cap
-q(\cup_{i=0}^{\frac{q+1}{4}}\mathbb{C}_{b})=\emptyset,\]
for $m+1\leq t\leq\frac{q-3}{4}$.\\

In the following, we proof
\[(\cup_{i=\frac{q+5}{4}}^
{\frac{q+1}{4}+m}\mathbb{C}_{b})\cap-q(\cup_{i=0}^{\frac{q+1}{4}}\mathbb{C}_{b})=\mathbb{C}_{\frac{(q-1)^2}{4}-m(q+1)},\]
i.e. $(\cup_{i=1}^
{m}\mathbb{C}_{s-(q+1)(i+\frac{q+1}{4})})\cap
-q(\cup_{i=0}^{\frac{q+1}{4}}\mathbb{C}_{s-(q+1)i})=\mathbb{C}_{\frac{(q-1)^2}{4}-m(q+1)}$. Then there exist two integers $l$ and $j$, where $1\leq l\leq m$, $0\leq j\leq \frac{q+1}{4}$, such that\\ \[s-(q+1)(l+\frac{q+1}{4})\equiv-q[s-(q+1)j]q^{2k} \bmod\ (q+1)n,\]
for some $k\in\{0,1\}$.

\begin{itemize}
	\item [(i)]  When $k=0$, we have
	$s-(q+1)(l+\frac{q+1}{4})\equiv-q[s-(q+1)j] \bmod\ (q+1)n$, which is equal to
    \begin{align}	
    \frac{q-3}{20}+q\cdot\frac{q-3}{10}\equiv l+qj \bmod\ n,
    \end{align}
	where $n=\frac{q^2+1}{5}=\frac{3q+1}{5}+q\cdot \frac{q-3}{5}$.
	From $1\leq l\leq m<\frac{3q+1}{5}$, we have the following results.
	\item [(a)] when $0\leq j\leq\frac{q-3}{5}$, $l+qj<n$.
	If equation (7) establish, we have $l=\frac{q-3}{20}=m$, $j=\frac{q-3}{10}=\frac{20m+3-3}{10}=2m$.
	\item [(b)]  when $\frac{q-3}{5}< j\leq\frac{q+1}{4}$, because $j$ is an integer, we have $\frac{q+2}{5}\leq j\leq\frac{q+1}{4}$, so $l+qj\geq qj\geq q\cdot\frac{q+2}{5}=\frac{q^2+2q}{5}>n$. From $ 1\leq l\leq m, \frac{q+2}{5}\leq j\leq\frac{q+1}{4}$, we can get
	$1+q\cdot\frac{q+2}{5}\leq l+qj\leq m+q\cdot\frac{q+1}{4}$, i.e.,
	$\frac{2q+4}{5}\leq l+qj-n\leq\frac{q^2+5q+20m-4}{20}<\frac{q-3}{20}+q\cdot\frac{q-3}{10}$, therefore equation
	$(7)$ is not establish.
\end{itemize}

\begin{itemize}
	\item [(ii)]  When $k=1$, we have
	$s-(q+1)(l+\frac{q+1}{4})\equiv-q^3[s-(q+1)j] \bmod\ (q+1)n$, which is equal to
    \begin{align}
    \frac{9q+3}{10}+q\cdot\frac{q-23}{20}\equiv j+ql \bmod\ n,
    \end{align} where
	$n=\frac{q^2+1}{5}=\frac{3q+1}{5}+q\cdot\frac{q-3}{5}$. From
	$0\leq j\leq\frac{q+1}{4}<\frac{3q+1}{5}$, $1\leq l\leq m<\frac{q-3}{5}=4m$,
	If equation (8) establish, we have $j=\frac{9q+3}{10}$, $l=\frac{q-23}{10}$. Because $\frac{9q+3}{10}>\frac{q+1}{4}$, it is a contradiction.
\end{itemize}

Next, we proof
\[(\cup_{i=\frac{q+5}{4}+m}^
{\frac{q+1}{4}+t}\mathbb{C}_{b})\cap
-q(\cup_{i=0}^{\frac{q+1}{4}}\mathbb{C}_{b})=\emptyset.
\]

If $(\cup_{i=\frac{q+5}{4}+m}^
{\frac{q+1}{4}+t}\mathbb{C}_{b})\cap
-q(\cup_{i=0}^{\frac{q+1}{4}}\mathbb{C}_{b})\neq\emptyset$,
for $m+1\leq t\leq\frac{q-3}{4}$, i.e.,\\
\[(\cup_{i=m+1}^
{t}\mathbb{C}_{s-(q+1)(i+\frac{q+1}{4})})\cap
-q(\cup_{i=0}^{\frac{q+1}{4}}\mathbb{C}_{s-(q+1)i})\neq\emptyset,\]
for  $m+1\leq t\leq\frac{q-3}{4}$, then there exist two integers $l$ and $j$, where $m+1\leq l\leq \frac{q-3}{4},~0\leq j\leq\frac{q+1}{4}$,
such that
\[s-(q+1)(l+\frac{q+1}{4})\equiv-q[s-(q+1)j]q^{2k} \bmod\ (q+1)n\]
 for some $k\in\{0,1\}$. We can seek contradictions as follows.

 \begin{itemize}
 	\item [(i)] When $k=0$, we have
 	$s-(q+1)(l+\frac{q+1}{4})\equiv-q[s-(q+1)j] \bmod\ (q+1)n$, which is equal to
 	\begin{align}
 	\frac{q-3}{20}+q\cdot\frac{q-3}{10}\equiv l+qj \bmod\ n,  \end{align}
    where $n=\frac{q^2+1}{5}=\frac{3q+1}{5}+q\cdot \frac{q-3}{5}$.
 	From $ m+1\leq l\leq\frac{q-3}{4}<\frac{3q+1}{5}$, we have the following results.
 	\item [(a)] when $0\leq j\leq\frac{q-3}{5}$, $l+qj<n$.
 	If equation (9) establish, we have $l=\frac{q-3}{20}$, $j=\frac{q-3}{10}$. Because $l_{min}=m+1>\frac{q-3}{20}=m$, it is a contradiction.
 	\item [(b)] when $\frac{q-3}{5}< j\leq\frac{q+1}{4}$, because $j$ is an integer, we have $\frac{q+2}{5}\leq j\leq\frac{q+1}{4}$, so $l+qj\geq qj\geq q\cdot\frac{q+2}{5}=\frac{q^2+2q}{5}>n$. From $m+1\leq l\leq\frac{q-3}{4}, \frac{q+2}{5}\leq j\leq\frac{q+1}{4}$,
 	we can get $m+1+q\cdot\frac{q+2}{5}\leq l+qj\leq\frac{q-3}{4}+q\cdot\frac{q+1}{4}$,
 	$m+1+\frac{q^2+2q}{5}\leq l+qj\leq\frac{q^2+2q-3}{4}$,
 	$m+\frac{2q+4}{5}\leq l+qj-n\leq\frac{q^2+10q-19}{20}<n$.
 	because
 	$\frac{q^2+10q-19}{20}<\frac{q-3}{20}+q\cdot\frac{q-3}{10}$, equation (9) is not establish.
 \end{itemize}

 \begin{itemize}
	\item [(ii)] When $k=1$, we have
	$s-(q+1)(l+\frac{q+1}{4})\equiv-q^3[s-(q+1)j] \bmod\ (q+1)n$, which is equal to
	\begin{align}
	\frac{9q+3}{10}+q\cdot\frac{q-23}{20}\equiv j+ql \bmod\ n,\end{align}
	where $n=\frac{q^2+1}{5}=\frac{3q+1}{5}+q\cdot\frac{q-3}{5}$.
    From $ 0\leq j\leq\frac{q+1}{4}<\frac{3q+1}{5}$, we have the following results.
	\item [(a)] when $m+1\leq l\leq\frac{q-3}{5}$, $l+qj<n$.
	If equation (10) establish, we have $j=\frac{9q+3}{10}$, $l=\frac{q-23}{20}$. Because $l_{min}=m+1>\frac{q-23}{20}=m-1$, it is a contradiction.
	\item [(b)] when $\frac{q-3}{5}< l\leq\frac{q-3}{4}$, because $l$ is an integer, $\frac{q+2}{5}\leq l\leq\frac{q-3}{4}$, so $j+ql\geq ql\geq q\cdot\frac{q+2}{5}=\frac{q^2+2q}{5}>n$. From $ \frac{q+2}{5}\leq l\leq\frac{q-3}{4}, 0\leq j\leq\frac{q+1}{4}$, we can get
	$q\cdot\frac{q+2}{5}\leq j+ql\leq\frac{q+1}{4}+q\cdot\frac{q-3}{4}$,
	i.e.,
	$\frac{q^2-1}{5}\leq j+ql-n\leq\frac{q^2-10q+1}{20}<n$, because
	$ \frac{9q+3}{10}+q\cdot\frac{q-23}{20}=\frac{q^2-5q+6}{20}>\frac{q^2-10q+1}{20}$, equation
	$(10)$ is not establish.
\end{itemize}

From the above discussions, we can see
\[(\cup_{i=\frac{q+5}{4}}^
{\frac{q+1}{4}+t}\mathbb{C}_{b})\cap
-q(\cup_{i=0}^{\frac{q+1}{4}}\mathbb{C}_{b})=\mathbb{C}_{\frac{(q-1)^2}{4}-m(q+1)},\]
for $m\leq t\leq\frac{q-3}{4}$.

Secondly, we show that
\[(\cup_{i=0}^
{\frac{q+1}{4}}\mathbb{C}_{b})\cap
-q(\cup_{i=\frac{q+5}{4}}^{\frac{q+1}{4}+t}\mathbb{C}_{b})=\mathbb{C}_{s-2m(q+1)}.\]
Since $-q\cdot[(\cup_{i=\frac{q+1}{4}+1}^
{\frac{q+1}{4}+t}\mathbb{C}_{b})\cap
-q(\cup_{i=0}^{\frac{q+1}{4}}\mathbb{C}_{b})]=-q\cdot\mathbb{C}_{\frac{(q-1)^2}{4}-m(q+1)}=\mathbb{C}_{s-2m(q+1)}$, \\
it follows that
\[(\cup_{i=0}^
{\frac{q+1}{4}}\mathbb{C}_{b})\cap
-q(\cup_{i=\frac{q+5}{4}}^{\frac{q+1}{4}+t}\mathbb{C}_{b})=\mathbb{C}_{s-2m(q+1)}.\]

Finally, we show that \[(\cup_{i=\frac{q+5}{4}}^{\frac{q+1}{4}+t}\mathbb{C}_{b})\cap-q(\cup_{i=\frac{q+5}{4}}^{\frac{q+1}{4}+t}\mathbb{C}_{b})=\emptyset,\]
for $m\leq t \leq \frac{q-3}{4}$. If\[(\cup_{i=\frac{q+5}{4}}^{\frac{q+1}{4}+t}\mathbb{C}_{b})\cap-q(\cup_{i=\frac{q+5}{4}}^{\frac{q+1}{4}+t}\mathbb{C}_{b})\neq\emptyset,\] i.e.,
\[(\cup_{i=1}^{t}\mathbb{C}_{s-(q+1)(i+\frac{q+1}{4})})\cap-q(\cup_{i=1}^{t}\mathbb{C}_{s-(q+1)(i+\frac{q+1}{4})})\neq\emptyset,\]
for $m\leq t \leq \frac{q-3}{4}$. Then there exist two integers $l$ and $j$, where $m\leq l,j \leq \frac{q-3}{4}$, such that
\[s-(q+1)(l+\frac{q+1}{4})\equiv-q[s-(q+1)(j+\frac{q+1}{4})]q^{2k} \bmod\ (q+1)n,\]
 for some $k\in\{0,1\}$. We can seek contradictions as follows.

 \begin{itemize}
 	\item [(i)]  When $k=0$, we have
 	$s-(q+1)(l+\frac{q+1}{4})\equiv-q[s-(q+1)(j+\frac{q+1}{4})] \bmod\ (q+1)n$, which is equal to
 	\begin{align}
 	\frac{13q+1}{20}+q\cdot\frac{q-23}{20}\equiv l+qj \bmod\ n,
 	\end{align}
 	where $n=\frac{q^2+1}{5}=\frac{3q+1}{5}+q\cdot \frac{q-3}{5}$. From
 	$ m\leq l\leq\frac{q-3}{4}<\frac{3q+1}{5}$, we have the following results.
 	\item [(a)]  when $m\leq j\leq\frac{q-3}{5}$, $l+qj<n$.
 	If equation (11) establish, we have $l=\frac{13q+1}{20}$, $j=\frac{q-23}{20}$. Because $j_{min}= m>\frac{q-23}{20}=m-1$, it is a contradiction.
 	\item [(b)] when $\frac{q-3}{5}< j\leq\frac{q-3}{4}$, because $j$ is an integer, we have $\frac{q+2}{5}\leq j\leq\frac{q-3}{4}$, so $l+qj\geq qj\geq q\cdot\frac{q+2}{5}=\frac{q^2+2q}{5}>n$. From $m\leq l\leq\frac{q-3}{4}, \frac{q+2}{5}\leq j\leq\frac{q-3}{4}$,
 	we can get $m+q\cdot\frac{q+2}{5}\leq l+qj\leq\frac{q-3}{4}+q\cdot\frac{q-3}{4}$, i.e.,
 	$m+\frac{2q-1}{5}\leq l+qj-n\leq\frac{q^2-10q-19}{20}<n$. because 
 	$\frac{q^2-10q+1}{20}>\frac{q^2-10q-19}{20}$,
 	equation (11) is not establish.
 \end{itemize}

 \begin{itemize}
	\item [(ii)] When $k=1$, we have
	$s-(q+1)(l+\frac{q+1}{4})\equiv-q^3[s-(q+1)(j+\frac{q+1}{4})] \bmod\ (q+1)n$,	which is equal to \begin{align}
	\frac{13q+1}{20}+q\cdot\frac{q-23}{20}\equiv j+ql \bmod\ n,
	\end{align}
	where $n=\frac{q^2+1}{5}=\frac{3q+1}{5}+q\cdot\frac{q-3}{5}$.
	From $ m\leq j\leq\frac{q-3}{4}<\frac{3q+1}{5}$, we have the following results.
	\item [(a)]  when $m\leq l\leq\frac{q-3}{5}$, $l+qj<n$.
	If equation (12) establish, we have $j=\frac{13q+1}{20}$, $l=\frac{q-23}{20}$. Because $l_{min}= m>l=\frac{q-23}{20}=m-1$, it is a contradiction.
	\item [(b)]  when $\frac{q-3}{5}< l\leq\frac{q-3}{4}$, because $l$ is an integer, we have $\frac{q+2}{5}\leq l\leq\frac{q-3}{4}$, so $l+qj\geq qj\geq q\cdot\frac{q+2}{5}=\frac{q^2+2q}{5}>n$. From $m\leq j\leq\frac{q-3}{4}, \frac{q+2}{5}\leq l\leq\frac{q-3}{4}$,
	we can get $m+q\cdot\frac{q+2}{5}\leq j+ql\leq\frac{q-3}{4}+q\cdot\frac{q-3}{4}$,
    i.e.,
	$m+\frac{2q-1}{5}\leq j+ql-n\leq\frac{q^2-10q-19}{20}$. Because
	$\frac{q^2-10q+1}{20}>\frac{q^2-10q-19}{20}$,
	equation (12) is not establish.
\end{itemize}

From Lemma 4.2, we have $c=4$. From Theorem 3.2, there exist entanglement assisted quantum codes with parameters $[[\frac{q^2+1}{5},\frac{q^2+1}{5}-q-4t+1,\frac{q+1}{2}+2t+2;4]]_q$, where $m\leq t\leq\frac{q-3}{4}$.\\

\noindent\textbf{Remark 4.8 } In Theorem 4.4, 4.5, and 4.7, $n+c-k= 2(d-1)$.
Then from Theorem $3.3$, the constructed EAQECCs attain entanglement-assisted quantum Singleton bound. Hence, these EAQECCs are optimal. 

Besides, when $q$ is an odd prime power with the form $20m+3$, our EAQECCs wirh parameters $[[\frac{q^2+1}{5},\frac{q^2+1}{5}-q-4t+1,\frac{q+1}{2}+2t+2;4]]_q$, where $m\leq t\leq\frac{q-3}{4}$, i.e., $12m+4\leq d=\frac{q+1}{2}+2t+2\leq 20m+4$ be even. Compared with  EAQECCs in \cite{ref25} with parmeters $[[\frac{q^2+1}{5},\frac{q^2+1}{5}-2d+6,d;4]]_q$, where $8m+3\leq d\leq12m+1$ be odd, in the case of the same $n$ and $c$, our EAQECCs have the larger $d$.  Compared with  EAQECCs
 in \cite{ref25} with parmeters $[[\frac{q^2+1}{5},\frac{q^2+1}{5}-2d+7,d;5]]_q$, where $16m+4\leq d\leq24m+4$ be even, in the case of the same $n$ and $d$, we use the lesser $c$ to attain the same net rate of EAQECCs.
 
 When $q$ is an odd prime power with the form $20m+7$, our EAQECCs wirh parameters $[[\frac{q^2+1}{5},\frac{q^2+1}{5}-q-4t+1,\frac{q+1}{2}+2t+2;4]]_q$, where $m\leq t\leq\frac{q-3}{4}$, i.e., $12m+6\leq d=\frac{q+1}{2}+2t+2\leq 20m+8$ be even. Compared with  EAQECCs in \cite{ref25} with parmeters $[[\frac{q^2+1}{5},\frac{q^2+1}{5}-2d+6,d;4]]_q$, where $16m+7\leq d\leq28m+11$ be odd, in the case of the same $n$ and $c$, both of EAQECCs are optimal, and our EAQECCs is new. Compared with  EAQECCs
 in \cite{ref25} with parmeters $[[\frac{q^2+1}{5},\frac{q^2+1}{5}-2d+7,d;5]]_q$, where $16m+8\leq d\leq24m+8$ be even, in the case of the same $n$ and $d$, we use the lesser $c$ to attain the same net rate of EAQECCs.\
  
 We give some examples in Table 1-5.\\

\begin{table}[htbp]
	\center \caption{Optimal EAQECCs from Theorem 4.5}
	\begin{tabular}{ccc}
		\toprule
		~~~~~~$q$ & ~~~~~~ $e$ &~~~~~~ $[[\frac{q^2+1}{5},\frac{q^2-6q+33}{5}-4t,\frac{3q-1}{5}+2t;4]]_q$ \\
		\midrule
		~~~~~~32 & ~~~~~~~5 & ~~~~~~$[[205,169,21;4]]_{32}$ \\
	    ~~~~~~32 & ~~~~~~~5 & ~~~~~~$[[205,165,23;4]]_{32}$ \\
	    ~~~~~~32 & ~~~~~~~5 & ~~~~~~$[[205,161,25;4]]_{32}$ \\
	    ~~~~~~32 & ~~~~~~~5 & ~~~~~~$[[205,157,27;4]]_{32}$ \\
	    ~~~~~~32 & ~~~~~~~5 & ~~~~~~$[[205,153,29;4]]_{32}$ \\
	    ~~~~~~32 & ~~~~~~~5 & ~~~~~~$[[205,149,31;4]]_{32}$ \\
	    ~~~~~~32 & ~~~~~~~5 & ~~~~~~$[[205,145,33;4]]_{32}$ \\
		\bottomrule
	\end{tabular}
\end{table}

\begin{table}[htbp]
	\center \caption{Optimal EAQECCs from Theorem 4.6}
	\begin{tabular}{ccc}
		\toprule
		~~~~~~$q$ & ~~~~~~ $e$ &~~~~~~ $[[\frac{q^2+1}{5},\frac{q^2-6q+29}{5}-4t,\frac{3q+1}{5}+2t;4]]_q$ \\
		\midrule
		~~~~~~8 & ~~~~~~~3 & ~~~~~~$[[13,5,7;4]]_{8}$ \\
		~~~~~~8 & ~~~~~~~3 & ~~~~~~$[[13,1,9;4]]_{8}$ \\
	
		\bottomrule
	\end{tabular}
\end{table}

\begin{table}[htbp]
	\center \caption{Optimal EAQECCs from Theorem 4.6}
	\begin{tabular}{ccc}
		\toprule
		~~~~~~$q$ & ~~~~~~ $e$ &~~~~~~ $[[\frac{q^2+1}{5},\frac{q^2-6q+29}{5}-4t,\frac{3q+1}{5}+2t;4]]_q$ \\
		\midrule
		~~~~~~128 & ~~~~~~~7 & ~~~~~	$[[3277,3125,~79;4]]_{128}$ \\
		~~~~~~128 & ~~~~~~~7 & ~~~~~~$[[3277,3121,~81;4]]_{128}$ \\
		~~~~~~128 & ~~~~~~~7 & ~~~~~~$[[3277,3117,~83;4]]_{128}$ \\
		~~~~~~128 & ~~~~~~~7 & ~~~~~~$[[3277,3113,~85;4]]_{128}$ \\
		~~~~~~128 & ~~~~~~~7 & ~~~~~~$[[3277,3109,~87;4]]_{128}$ \\
		~~~~~~128 & ~~~~~~~7 & ~~~~~~$[[3277,3105,~89;4]]_{128}$ \\
		~~~~~~128 & ~~~~~~~7 & ~~~~~~$[[3277,3101,~91;4]]_{128}$ \\
		~~~~~~128 & ~~~~~~~7 & ~~~~~~$[[3277,3097,~93;4]]_{128}$ \\
		~~~~~~128 & ~~~~~~~7 & ~~~~~~$[[3277,3093,~95;4]]_{128}$ \\
		~~~~~~128 & ~~~~~~~7 & ~~~~~~$[[3277,3089,~97;4]]_{128}$ \\
		~~~~~~128 & ~~~~~~~7 & ~~~~~~$[[3277,3085,~99;4]]_{128}$ \\
		~~~~~~128 & ~~~~~~~7 & ~~~~~~$[[3277,3081,101;4]]_{128}$  \\
		~~~~~~128 & ~~~~~~~7 & ~~~~~~$[[3277,3077,103;4]]_{128}$ \\
		~~~~~~128 & ~~~~~~~7 & ~~~~~~$[[3277,3073,105;4]]_{128}$ \\
		~~~~~~128 & ~~~~~~~7 & ~~~~~~$[[3277,3069,107;4]]_{128}$ \\
		~~~~~~128 & ~~~~~~~7 & ~~~~~~$[[3277,3065,109;4]]_{128}$ \\
		~~~~~~128 & ~~~~~~~7 & ~~~~~~$[[3277,3061,111;4]]_{128}$ \\
		~~~~~~128 & ~~~~~~~7 & ~~~~~~$[[3277,3057,113;4]]_{128}$ \\
		~~~~~~128 & ~~~~~~~7 & ~~~~~~$[[3277,3053,115;4]]_{128}$ \\
		~~~~~~128 & ~~~~~~~7 & ~~~~~~$[[3277,3049,117;4]]_{128}$ \\
		~~~~~~128 & ~~~~~~~7 & ~~~~~~$[[3277,3045,119;4]]_{128}$ \\
		~~~~~~128 & ~~~~~~~7 & ~~~~~~$[[3277,3041,121;4]]_{128}$ \\
		~~~~~~128 & ~~~~~~~7 & ~~~~~~$[[3277,3037,123;4]]_{128}$ \\
		~~~~~~128 & ~~~~~~~7 & ~~~~~~$[[3277,3033,125;4]]_{128}$ \\
		~~~~~~128 & ~~~~~~~7 & ~~~~~~$[[3277,3029,127;4]]_{128}$ \\
		~~~~~~128 & ~~~~~~~7 & ~~~~~ $[[3277,3025,129;4]]_{128}$  \\
		\bottomrule
	\end{tabular}
\end{table}

\begin{table}[htbp]
		\center \caption{Optimal EAQECCs from Theorem 4.7}
	\begin{tabular}{ccc}
		\toprule
		~~~~~~$q$ & ~~~~~~ $m$ &~~~~~~ $[[\frac{q^2+1}{5},\frac{q^2+1}{5}-q-4t+1,\frac{q+1}{2}+2t+2;4]]_q$ \\
		\midrule
		~~~~~~23 & ~~~~~~~1 & ~~~~~~$[[106,80,16;4]]_{23}$ \\
		~~~~~~23 & ~~~~~~~1 & ~~~~~~$[[106,76,18;4]]_{23}$ \\
		~~~~~~23 & ~~~~~~~1 & ~~~~~~$[[106,72,20;4]]_{23}$  \\
		~~~~~~23 & ~~~~~~~1 & ~~~~~~$[[106,68,22;4]]_{23}$ \\
	    ~~~~~~23 & ~~~~~~~1 & ~~~~~~$[[106,64,24;4]]_{23}$ \\
		\bottomrule
	\end{tabular}
\end{table}

\begin{table}[htbp]
	\center \caption{Optimal EAQECCs from Theorem 4.7}
	\begin{tabular}{ccc}
		\toprule
		~~~~~~$q$ & ~~~~~~ $m$ &~~~~~~ $[[\frac{q^2+1}{5},\frac{q^2+1}{5}-q-4t+1,\frac{q+1}{2}+2t+2;4]]_q$ \\
		\midrule
		~~~~~~47 & ~~~~~~~2 & ~~~~~~$[[442,388,30;4]]_{47}$ \\
		~~~~~~47 & ~~~~~~~2 & ~~~~~~$[[442,384,32;4]]_{47}$ \\
		~~~~~~47 & ~~~~~~~2 & ~~~~~~$[[442,380,34;4]]_{47}$ \\
		~~~~~~47 & ~~~~~~~2 & ~~~~~~$[[442,376,36;4]]_{47}$ \\
		~~~~~~47 & ~~~~~~~2 & ~~~~~~$[[442,372,38;4]]_{47}$ \\
		~~~~~~47 & ~~~~~~~2 & ~~~~~~$[[442,368,40;4]]_{47}$ \\
		~~~~~~47 & ~~~~~~~2 & ~~~~~~$[[442,364,42;4]]_{47}$ \\
		~~~~~~47 & ~~~~~~~2 & ~~~~~~$[[442,360,44;4]]_{47}$ \\
		~~~~~~47 & ~~~~~~~2 & ~~~~~~$[[442,356,46;4]]_{47}$ \\
		~~~~~~47 & ~~~~~~~2 & ~~~~~~$[[442,352,48;4]]_{47}$ \\
		\bottomrule
	\end{tabular}
\end{table}

In the above part of this section, we have discussed three families of entanglement-assisted quantum MDS codes constructed from constacyclic codes. In the following part
of this section, we will find that there exist EAQECCs with
maximal entanglement.\\

\noindent\textbf{Theorem 4.9 }\emph{Let $q=2^{e}$ with
	$e\equiv1(\bmod\ 4)$. Let $n=\frac{q^2+1}{5}$, $s=\frac{(q+6)n}{2}$ and
	$r=\frac{q^2-q}{2}$, where $r=s-\frac{(q+1)(n+1)}{2}$. If $\mathcal C$ is a $q^2$-ary constacyclic code of length $n$
	with defining set $Z=\mathbb{C}_\frac{q^2-q+3}{5}\cup\mathbb{C}_\frac{2q^2-2q+1}{5}$,
	then there exsit maximal-entanglement entanglement-assisteed quantum codes with parameters $[[\frac{q^2+1}{5},\frac{q^2+1}{5}-4,d\geq2;4]]_q$.}\\

\noindent\textbf{Proof.} Assume that the defining set of the constacyclic code $\mathcal C$ is $Z=\mathbb{C}_\frac{q^2-q+3}{5}\cup\mathbb{C}_\frac{2q^2-2q+1}
{5}$, then $\mathcal C$ is a constacyclic code with parameters $[\frac{q^2+1}{5},\frac{q^2+1}{5}-4,d\geq2]_{q^2}$ from Theorem $2.1$ and Lemma $3.1$. Since $Z\cap(-qZ)=\mathbb{C}_\frac{q^2-q+3}{5}\cup\mathbb{C}_\frac{2q^2-2q+1}
{5}$, it follows that $c=4$ from Lemma 4.2. Therefore,there exist maximal-entanglement entanglement-assisted quantum codes with parameters $[[\frac{q^2+1}{5},\frac{q^2+1}{5}-4,d\geq2;4]]_q$.\\

Similar to the proof of Theorem 4.9, we can get another two theorems as follows. \\

\noindent\textbf{Theorem 4.10 }\emph{Let $q=2^{e}$ with
	$e\equiv3(\bmod\ 4)$. Let $n=\frac{q^2+1}{5}$, $s=\frac{(q+6)n}{2}$ and
	$r=\frac{q^2-q}{2}$,  where $r=s-\frac{(q+1)(n+1)}{2}$. If $\mathcal C$ is an $q^2$-ary constacyclic code of length $n$
	with defining set $Z=\mathbb{C}_\frac{q^2-2q+2}{5}\cup\mathbb{C}_\frac{3q^2-q+1}{5}$, then there exist maximal-entanglement entanglement-assisteed quantum codes with parameters $[[\frac{q^2+1}{5},\frac{q^2+1}{5}-4,d\geq2;4]]_q$.}\\

\noindent\textbf{Theorem 4.11 }\emph{Let $n=\frac{q^2+1}{5}$
	and $s=\frac{q^2+1}{2}$, where $q$ is an odd prime power with the form $20m+3$ or $20m+7$ and $m$ is a positive integer. If $\mathcal C$ is a $q^2$-ary constacyclic code of length $n$ with defining set  $Z=\mathbb{C}_{\frac{(q-1)^2}{4}-m(q+1)}\cup\mathbb{C}
	_{s-2m(q+1)}$ or $Z=\mathbb{C}_{\frac{(q-1)^2}{4}-m(q+1)}\cup\mathbb{C}
	_{s-(2m+1)(q+1)}$, then there exist maximal-entanglement entanglement-assisteed quantum codes with parameters $[[\frac{q^2+1}{5},\frac{q^2+1}{5}-4,d\geq2;4]]_q$.}\\

\dse{5~~Conclusion} In this paper, we have constructed three classes
of optimal EAQECCs and three classes of maximal-entanglement entanglement-assisted quantum codes from constacyclic codes over the finite field
$\mathbb{F}_{q^2}$ of length $n=\frac{q^2+1}{5}$, where $q$ is some prime power. The construction is through cyclotomic
cosets and ideal theory. According to the  entanglement-assisted quantum
Singleton bound, the resulting entanglement-assisted quantum
codes are optimal and different from
the codes available in the literature. It would be interesting to construct optimal EAQECCs
 from other types of constacyclic codes.

\end{document}